\documentclass[fleqn,usenatbib]{mnras}

\usepackage{newtxtext,newtxmath}
\usepackage{ulem}
\usepackage{physics}

\usepackage[T1]{fontenc}

\DeclareRobustCommand{\VAN}[3]{#2}
\let\VANthebibliography\thebibliography
\def\thebibliography{\DeclareRobustCommand{\VAN}[3]{##3}\VANthebibliography}

\usepackage{graphicx}	
\usepackage{amsmath}	

\title[Energy Dissipation of Magnetic Bullets]{Dynamical Energy Dissipation of  Relativistic Magnetic Bullets}

\author[Y. Kusafuka et al.]{
Yo Kusafuka,$^{1}$\thanks{E-mail: kusafuka@icrr.u-tokyo.ac.jp}
Katsuaki Asano,$^{1}$
Takumi Ohmura$^{1}$
Tomohisa Kawashima$^{1}$
\\
$^{1}$Institute for Cosmic Ray Research, The University of Tokyo, 5-1-5 Kashiwanoha, Kashiwa, Chiba 277-8582, Japan
}

\date{Accepted XXX. Received YYY; in original form ZZZ}

\pubyear{2023}

\begin{document}
\label{firstpage}
\pagerange{\pageref{firstpage}--\pageref{lastpage}}
\maketitle

\begin{abstract}
To demonstrate the magnetic energy dissipation via relativistic shocks, we carry out spherically symmetrical one-dimensional special relativistic magneto-hydrodynamic simulations of highly magnetised outflows with an adaptive mesh refinement method. We first investigate the detail of the dynamical energy dissipation via interaction between a single ejecta and an external medium. The energy dissipation timescales, which affect the early behaviour of the afterglow emission in gamma-ray bursts, are estimated for a wide range of magnetisation. In addition, we demonstrate the internal shock dissipation in multiple interactions between magnetically dominated relativistic ejecta and kinetically dominated non-relativistic winds. Our numerical results show that $\sim 10$\% of the magnetic energy in the ejecta can be converted into the thermal energy of the relativistic and low-magnetised outflows via shocks in the rarefaction waves or the winds. Such hot and less magnetised outflows are relevant for observed non-thermal emissions in blazars or gamma-ray bursts. 
\end{abstract}




\section{Introduction}

Relativistic outflows or jets with more than 99\% of the light speed emerge in pulsar wind nebulae, gamma-ray bursts (GRBs), and active galactic nuclei (AGNs). Such relativistic jets are thought to be launched through magnetic processes \citep{1977MNRAS.179..433B}, which implies magnetically dominated outflows of $\sigma\gg1$, where $\sigma$ is the magnetisation parameter defined as the ratio of the magnetic energy to the rest mass energy. Dissipation of the energy through collision with interstellar mediums (ISM) or stellar winds causes multi-wavelength radiation. This external shock model can explain GRB afterglows \citep{1992MNRAS.258P..41R,1995ApJ...455L.143S,1997ApJ...476..232M}. The energy dissipation of jets can also occur inside themselves due to their variability. This internal shock model is often used to explain the prompt emission of GRBs and blazar flares \citep{1978MNRAS.184P..61R,1994ApJ...430L..93R,1997ApJ...490...92K,1998MNRAS.296..275D}.

Shock waves have been thought to be vital in the radiation from relativistic jets. Shock waves can dissipate the kinetic energy of jets, a part of which can be transferred to the energy of non-thermal particles via the diffusive shock acceleration \citep{1949PhRv...75.1169F,1978ApJ...221L..29B,1978MNRAS.182..147B}. 
However, the dissipation efficiency of the internal shocks for magnetically dominated outflows is quite low \citep{1984ApJ...283..694K} compared to observationally implied radiative efficiency of GRBs and Blazars \citep{2012Sci...338.1445N}. In addition, several theoretical arguments suggest that the diffusive shock acceleration is inefficient for $\sigma>10^{-3}$ \citep{2011ApJ...726...75S,2013ApJ...771...54S,2018MNRAS.477.5238P}. Therefore, conventional internal shock models have difficulty in explaining particle acceleration and energy dissipation for magnetically dominated relativistic jets.

Magnetic reconnection is another candidate process to accelerate particles efficiently in strongly magnetised plasma \citep{2011ApJ...726...75S,PhysRevLett.113.155005,2014ApJ...783L..21S,2015MNRAS.450..183S,2017ApJ...843L..27W,2019ApJ...880...37P}. 
Magnetic reconnection can dissipate magnetic energy efficiently \citep{2015MNRAS.450..183S} and lead to ultra-fast variability of radiation \citep{2009MNRAS.395L..29G,2016MNRAS.462.3325P}, which can explain the observed few minutes timescale of gamma-ray variability for blazars \citep{2007ApJ...664L..71A,2007ApJ...669..862A,2013ApJ...762...92A,2018ApJ...856...95A,2022A&A...668A.152P}. 
However, gamma-ray flares of some BL Lac objects are considered to be emitted from kinetically dominated jets \citep{2011ApJ...736..131A,2011ApJ...738...25A,2012A&A...542A.100A,2015MNRAS.451..739A,2010MNRAS.401.1570T,2016MNRAS.456.2374T,2019MNRAS.484.1192S,2020A&A...640A.132M}, which is not favourable for magnetic reconnection models \citep{2015MNRAS.450..183S}. 

The Imaging X-ray Polarimetry Explorer \citep{2022JATIS...8b6002W} observation detected X-ray polarisation from BL Lac objects Mrk 501 \citep{2022ApJ...938L...7D} and Mrk 421 \citep{2022Natur.611..677L}. These results show high X-ray polarisation degree and electric vector position angle almost along the jet direction without short time variability and polarisation angle swing. If magnetic reconnection occurs in the magnetic turbulence, significant polarisation variability and polarisation angle swing should be observed \citep{2018ApJ...862L..25Z,2021MNRAS.501.2836B}. On the other hand, a shock acceleration model \citep{2021Galax...9...37T} is consistent with all the observed features.

In the external shock models, the energy of the magnetised ejecta is efficiently transferred to the external medium. Thus, the dynamical energy dissipation by inducing shock waves in a weakly magnetised medium is the easiest method to transfer the magnetic energy into radiation. Even in the internal shock models, the intermittent ejection of jets may be one of the possible solutions for realising the energy dissipation of highly magnetised outflows.
The highly variable emission in blazers and GRBs is attributed to the variable activity of the central engine. Some numerical studies also show a short timescale intermittent ejection of jets \citep{2019MNRAS.490.4811C,2021MNRAS.508.1241C}. Assuming intermittent ejections, we can imagine that low-magnetised mediums can invade between intermittent jets. The interaction of the highly magnetised outflow and weakly magnetised medium induces the dissipation of the magnetic energy into low magnetised regions \citep{2011MNRAS.411.1323G,2012MNRAS.422..326K}. Possible mediums in the internal shock models are winds from accretion disks or rarefaction waves propagating from the preceding magnetised ejecta as discussed in \citet{2012MNRAS.422..326K}.

In this paper, we demonstrate how magnetic energy initially possessed by relativistic ejecta (magnetic bullet) is converted into thermal energy in low-magnetised plasma by performing one-dimensional (1D) special relativistic magneto-hydrodynamical (SRMHD) simulations. As the IXPE results for Mrk 421 and Mrk 501 suggest, we focus on magnetic energy dissipation without magnetic reconnection.
In our 1D fluid systems, the kinetic or magnetic energy is dissipated into the thermal energy, which may implicitly include the energy of non-thermal particles produced via plasma-kinematic effects such as stochastic shock acceleration, or the turbulence energy in realistic multi-dimensional systems. Such turbulence is also responsible for the production of non-thermal particles \citep[see e.g.][and references therein]{2019ApJ...877...71T}.
A part of the dissipated energy in our simulations can be released as radiation from such non-thermal particles accelerated by shocks or turbulence \citep[e.g.][]{2009ApJ...705.1714A,2014ApJ...780...64A}.

The paper is organised as follows. In Section \ref{sec:sim}, we describe the fundamental formulae and our numerical scheme. In Section \ref{sec:GRB}, we show the results of interactions between a single magnetised ejecta and interstellar medium to explain the detail of the magnetic energy conversion mechanism. In Section \ref{sec:blazar}, we show the results of interactions between numerous magnetic bullets and weakly magnetised winds to estimate the average dissipation efficiency. The conclusion is summarised in Section \ref{sec:dis}.

\section{Simulation method}\label{sec:sim}

In this paper, we consider magnetically-dominated relativistic ejecta (magnetic bullets) launched from a central engine. To investigate the energy dissipation of magnetic bullets via interaction with external medium or non-relativistic winds, we carry out 1D SRMHD simulations assuming spherically symmetric geometry. 1D simulations are free from fluid instabilities such as Rayliegh-Taylor, Kelvin-Helmholtz, and Kink instabilities, which are also candidate mechanisms to dissipate magnetic energy by reconnection. Besides, microscopic instability like Weibel instability can amplify the magnetic energy. We neglect the effects of such instabilities in this paper. We also neglect the effects of the equatorial motion of the surrounding medium, which may lead to modifying the magnetic field structures of jets by recollimation shocks or turbulences at the outer radii. Throughout this paper, we focus on a magnetic energy dissipation mechanism without the above multi-dimensional effects.

In the spherical coordinate, we consider only the radial component for the fluid velocity $\vb*{v}=(v,0,0)$ with magnetic field perpendicular to the direction of the velocity ($\theta$-component) $\vb*{B}=(0,B,0)$, both are measured at the rest frame of the external medium. Because our interest is a flow just around the jet axis, the spherically symmetric approximation is adopted as a local geometry around the jet axis. Whereas, the mass density $\rho$, the gas pressure $p$, and the energy density $\epsilon$ are measured at the fluid co-moving frame. The equation of state is
\begin{equation}
\epsilon=\frac{p}{\hat{\gamma}-1}+\rho c^2,
\label{eq:EoS}
\end{equation}
where $\hat{\gamma}$ is the adiabatic index, which can be approximated as \citep{2007MNRAS.378.1118M}
\begin{equation}
\hat{\gamma}=1+\frac{\epsilon+\rho c^2}{3\epsilon}.
\label{eq:index}
\end{equation}
The mass, energy, momentum conservation laws, and the induction equation are described as \citep[see e.g.][]{2012ApJS..198....7M}
\begin{equation}
\frac{1}{c}\frac{\partial \rho\Gamma}{\partial t}+\frac{1}{r^2}\frac{\partial}{\partial r}\left(r^2\rho \Gamma \beta\right)=0,
\end{equation}
\begin{equation}
\begin{split}
\frac{1}{c}\frac{\partial}{\partial t}\left[\left(\epsilon+p+\right .\right.&\left.\left.\frac{B^2}{4\pi\Gamma^2}\right)\Gamma^2-p-\frac{B^2}{8\pi\Gamma^2}\right]\\
&+\frac{1}{r^2}\frac{\partial}{\partial r}\left(r^2 \left[\left(\epsilon+p+\frac{B^2}{4\pi\Gamma^2}\right)\Gamma^2\beta \right]\right)=0,
\end{split}
\label{eq:energy}
\end{equation}
\begin{equation}
\begin{split}
\frac{1}{c}\frac{\partial}{\partial t}&\left[\left(\epsilon+p+\frac{B^2}{4\pi\Gamma^2}\right)\Gamma^2\beta\right]\\
&+\frac{1}{r^2}\frac{\partial}{\partial r}\left(r^2\left[\left(\epsilon+p+\frac{B^2}{4\pi\Gamma^2}\right)\Gamma^2\beta^2+p+\frac{B^2}{8\pi\Gamma^2} \right] \right)=\frac{2p}{r},
\end{split}
\end{equation}
\begin{equation}
\frac{1}{c}\frac{\partial B}{\partial t}+\frac{1}{r}\frac{\partial }{\partial r}\left(r\beta B \right)=0,
\end{equation}
respectively, where $\beta=v/c$ is the fluid velocity normalised by the speed of light, and $\Gamma=1/\sqrt{1-\beta^2}$ is the Lorentz factor of the fluid. For the parameterisation in our simulations, we define the magnetisation parameter $\sigma$ as
\begin{equation}
    \sigma\equiv\frac{B^2}{4\pi(\epsilon+p)\Gamma^2}.
\end{equation}

For the spatial interpolation of variables, we adopt the 2nd-order MUSCL scheme \citep{1979JCoPh..32..101V}. For the time interpolation of variables, we use the 2nd-order Runge-Kutta method and set the CFL number around 0.1. We use the minmod function as a flux limiter \citep{1986AnRFM..18..337R}, and compute numerical flux by approximate Riemann solver, the CENTRAL scheme \citep{KURGANOV2000241}, or so-called Rusanov scheme \citep{RUSANOV1962304}. For the primitive recovery, we use the Newton-Rhapson method \citep{2006MNRAS.368.1040M}. Our code successfully passed some numerical tests including simple advection tests \citep[e.g.][]{1988ApJ...332..659E,1992ApJ...388..415S} and several shock tube tests given in \citet{2006JFM...562..223G}. Although these numerical tests are the case for weakly magnetised ($\sigma\ll1$) and mildly relativistic ($\Gamma\sim1$), our code can treat the dynamics of strongly magnetised ($\sigma\gg1$) relativistic ($\Gamma\gg1$) shock waves as shown in the following Section \ref{sec:GRB}. These results agree with the analytic formulae for relativistic shock waves, which guarantees the robustness of our code for $\Gamma\gg1$ and $\sigma\gg1$.

Relativistic ($\Gamma\gg1$) magnetic bullets with a high $\sigma>1$ lead to strong shock structures in our simulations. To resolve such shocks and avoid numerical dissipation, an ultra-high resolution is required \citep{2009A&A...494..879M}. We implement the adaptive mesh refinement \citep{1984JCoPh..53..484B} to obtain a higher resolution around all discontinuities of magnetic pressure (AMR, see Figure \ref{fig:AMR} in Appendix \ref{Appendix A}). Our implemented AMR successfully suppress the numerical errors with more than 100 times less computational costs. Even with our high-resolution scheme, we cannot avoid a slight numerical dissipation of magnetic energy. We, however, have confirmed that the numerical dissipation is negligible for our purpose by resolution studies for the energy dissipation efficiency. We use the MPI parallelisation approach to reduce computational time. It takes a few days to finish the simulations in Section \ref{sec:GRB} and a few tens of days in Section \ref{sec:blazar} using 200 CPU cores with MPI parallelisation.

\section{Single magnetic bullet} \label{sec:GRB}

In this section, we study a single relativistic ejecta plunging into a homogeneous low-magnetised external medium. The interaction between the ejecta and the medium generates strong shocks. The initial magnetic energy of the ejecta is converted into the kinetic and thermal energy of the shocked external medium. This energy transfer is the same mechanism that occurred in GRB afterglows \citep{1992MNRAS.258P..41R,1995ApJ...455L.143S,1997ApJ...476..232M}. 

Our purpose in this section is a detailed study of the magnetic energy conversion mechanism via external shock formation. The final stage for the forward shock should follow the well-known Blandford-McKee solution \citep{1976PhFl...19.1130B}. The reverse shock crossing time becomes shorter as the magnetisation increases \citep{2005ApJ...628..315Z}, and the initial ejecta energy is almost transferred to the shocked external medium \citep{2009A&A...494..879M}. The simulations in this section are also test calculations by comparing the results with the results in \citet{2009A&A...494..879M}. While \citet{2009A&A...494..879M} adopts $\Gamma=15$ and $\sigma\leqq1$, we simulate with a larger $\sigma$ (up to 10 in magnetic bullet) and a comparable Lorentz factor $\Gamma=10$. \footnote{The animations are available at \url{https://www.youtube.com/playlist?list=PLgnUM4yGp9oLxqXlcKcc8ftJne9kTpHEs}}

\subsection{Initial setup}

The simulation box is in the external medium rest frame. For the external medium, we assume that the magnetic field and the number density are homogeneous and set as $1\ \mu$G and $n_0=1\ \rm{cm}^{-3}$, respectively. The temperature is also fixed as 1 MeV to stabilise the simulations, but it is sufficiently cold compared to proton rest mass energy.

From the definition of the magnetisation $\sigma$, we can write the initial magnetic field of the ejecta as
\begin{equation}
    B=\sqrt{4\pi(\epsilon+p)\Gamma^2_0\sigma_0},
\end{equation}
where $\sigma_0$ and $\Gamma_0=10$ are the initial magnetisation and the Lorentz factor of the ejecta, respectively. We define the deceleration radius $R_{\rm dec}$ as \citep{1995ApJ...455L.143S}
\begin{equation}
    R_{\rm dec}=\left(\frac{3E_0}{4\pi n_0m_pc^2\Gamma_0^2}\right)^{1/3}.
    \label{eq:rdec}
\end{equation}
At this radius, the initial ejecta energy $E_0$ is expected to be converted into the energy of the shocked external medium. The inner edge of the ejecta is initially set at $R_0=0.11R_{\rm dec}$. We fix the initial width of the ejecta as $\Delta=R_{\rm dec}/\Gamma^2_0$. With this setup, the initial width $\Delta=0.01R_{\rm dec}$ is thicker than $R_0/\Gamma_0^2=0.01R_0$ (thick shell model), which can be realised by a significantly longer duration of the central engine activity. In our simulations, we consider only the thick shell model because the thin shell model requires a much better spatial resolution. We assume that the ejecta has constant magnetisation $\sigma_0$, temperature $T_0=100$ MeV, and Lorentz factor $\Gamma_0$, but the mass density follows as $\rho=\rho_0(r/R_0)^{-2}$ in the initial condition. The initial energy of the ejecta is set to be $E_0=10^{50}$ erg, which is calculated with
\begin{equation}
    E_0=\int_{R_0}^{R_0+\Delta}4\pi r^2dr\left[(1+\sigma_0)(\epsilon+p)\Gamma_0^2\right].
\end{equation}
Using Eqs.(\ref{eq:EoS}) and (\ref{eq:index}), we can write the normalisation of the mass density $\rho_0$ as
\begin{equation}
    \rho_0=\frac{E_0m_pc^2R_0^2}{2\pi\left(5T_0+\sqrt{9T_0^2+4(m_pc^2)^2}\right)(1+\sigma_0)\Gamma^2_0\Delta}.
    \label{eq:rho}
\end{equation}
The deceleration time $t_{\rm dec}\simeq (R_{\rm dec}-R_0)/c$ is $1.6\times10^6$ s in our setup.

The simulation box size is $[0.1R_{\rm dec},5R_{\rm dec}]$. The outer boundary is set to be an open boundary. The inner boundary is set to be an injection boundary, where we inject a plasma with a significantly low luminosity with $\Gamma=10$ continuously. This injection can prevent the regions just behind the ejecta from becoming too dilute (avoiding a numerical floor). The number of static cells is $5\times10^5$, effectively $8\times10^6$ via the AMR procedure (see Appendix \ref{Appendix A}). The initial width of the ejecta is resolved by almost $10^3$ static cells and effectively almost $10^4$ static cells. 

\subsection{Structure of the shock waves\label{subsec:shock}}

\begin{figure}
\includegraphics[width=\columnwidth]{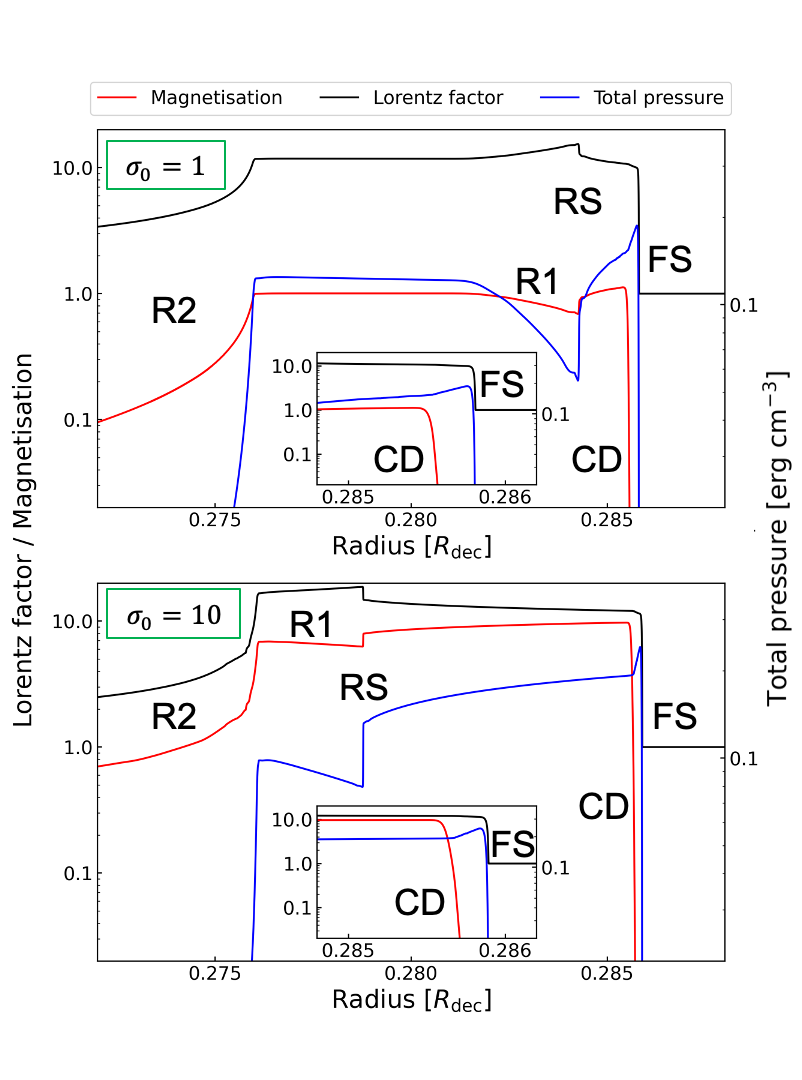}
\caption{Snapshots of the fluid-structure in the external shock models at $3\times10^5$ s. The red, black, and blue lines show $\sigma$, $\Gamma$, and the total pressure. The top panel is for the model with $\sigma_0=1$, while the bottom panel is for the magnetic bullet model ($\sigma_0=10$). The forward shock (FS) appears at the discontinuity of $\Gamma$ located around $0.285R_{\rm dec}$. The contact discontinuity (CD) is seen at the discontinuity of $\sigma$ just behind the FS. The reverse shock (RS) is the discontinuities of both $\Gamma$ and $\sigma$ located at 0.278 for $\sigma_0=10$ (0.284 for $\sigma_0=1$). The rarefaction waves (R1, R2) lead to pressure gradients in the ejecta. The insets are close-ups around the FS and CD.
\label{fig:CD}}
\end{figure}

Just after the start of the calculation, two rarefaction waves start to propagate from the front and the rear edges of the ejecta, respectively. As the rarefaction waves propagate, pressure gradients are formed inside the ejecta. One of them accelerates the front edge of the ejecta, while the other decelerates the rear edge of it \citep{2011MNRAS.411.1323G}. The rarefaction wave accelerates the ejecta faster for higher $\sigma_0$. As the forward shock propagates into the external medium, the increased thermal pressure of the shocked external medium generates the reverse shock to decelerate the ejecta.

Figure \ref{fig:CD} shows snapshots of $\Gamma$, $\sigma$, and the total pressure around the shocks at $t=3\times10^5$ s. The mass density profiles are not shown for simplicity. From right to left, we can see the forward shock (FS), where $\Gamma$ is jumping, the contact discontinuity (CD), where $\sigma$ is jumping, the reverse shock (RS), where $\Gamma$ and $\sigma$ are jumping again, the rarefaction wave accelerating the ejecta (R1), and the rarefaction wave decelerating the ejecta (R2), which forms the tail of the ejecta. Because the propagation speed of the magneto-sonic waves depends on magnetisation, the positions of R1 and RS are sensitive to $\sigma_0$. In Figure \ref{fig:CD}, for $\sigma_0=1$, the wave R1 has not yet reached the rear edge of the ejecta so that the initial flat structures of $\Gamma$ and $\sigma$ remain at 0.276--0.281 $R_{\rm dec}$, and the RS position is still close to the head of the ejecta. For $\sigma_0=10$, the wave R1 has already reached the rear edge of the ejecta, which eliminates the initial flat structure, and the RS position is close to the rear edge of the ejecta.

\subsection{Behaviour of the reverse shock \label{subsec:RS}}

\begin{figure}
\includegraphics[width=\columnwidth]{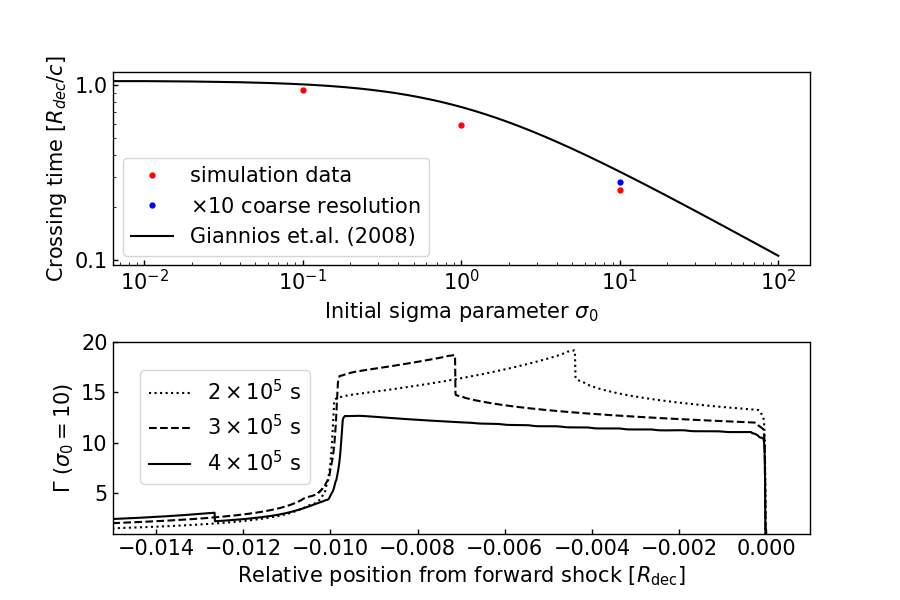}
\caption{Top: The reverse shock crossing timescale as a function of the initial magnetisation. The normalisation of analytic curve \citep{2008A&A...478..747G} is determined by our simulation data of $\sigma_0=0$. Bottom: The Lorentz factor distribution around the reverse shock at different times. 
\label{fig:RS}}
\end{figure}

The reverse shock propagates into the ejecta. We define the time at which the reverse shock crosses the ejecta as $t_\Delta$. In a hydrodynamic case ($\sigma_0=0$), \citet{1995ApJ...455L.143S} provided an analytical formula of $t_\Delta$ as
\begin{equation}
    t_\Delta(\sigma=0)=\Gamma_0^{1/2}\left(R_{\rm dec}\right)^{3/4}\Delta^{1/4},
\end{equation}
where $\Delta$ is the shell thickness in the external medium rest frame. For arbitrary magnetised cases, \citet{2005ApJ...628..315Z} estimated the reverse shock crossing time $t_\Delta(\sigma)$ as a function of $\sigma$, which is approximated as \citep{2008A&A...478..747G}
\begin{equation}
\label{RS}
    t_\Delta(\sigma)\simeq t_\Delta(0)(1+\sigma)^{-1/2}.
\end{equation}
As shown in Figure \ref{fig:RS}, the reverse shock crossing time $t_\Delta$ well agrees within $\sim$ 10\% with Eq. (\ref{RS}), where the normalisation $t_\Delta(0)$ is given by our simulation data of $\sigma_0=0$. A simulation with 10 times coarse resolution shows that the crossing time $t_\Delta$ becomes slightly longer. This is due to the numerical expansion of the edge of the ejecta due to the insufficient resolution (see Appendix \ref{Appendix A}). In GRB afterglows, the reverse shock crossing time corresponds to the time of the peak flux from the reverse shock \citep{2005ApJ...628..315Z}. From our simulations, the early peak time of the reverse shock emission for high-$\sigma$ ejecta is confirmed.

\subsection{Behaviour of the forward shock\label{subsec:FS}}

\begin{figure}
\includegraphics[width=\columnwidth]{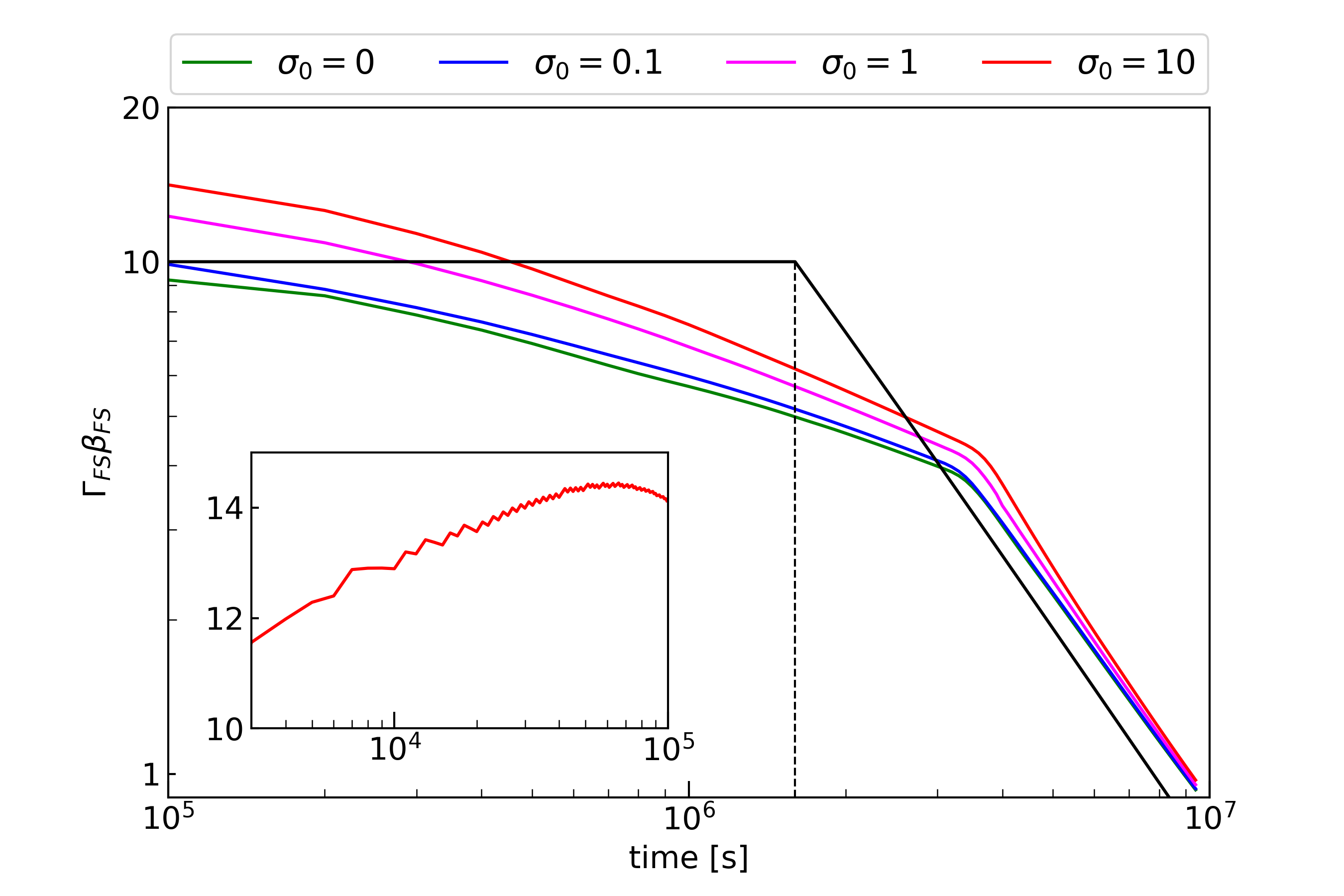}
\caption{The time evolution of $\Gamma\beta$ of the shocked external medium just behind the forward shock. The vertical dashed line shows the deceleration time $t_{\rm dec}$ defined with Eq. (\ref{eq:rdec}). Each coloured line corresponds to a different initial magnetisation $\sigma_0$. The black line shows the frequently used approximation. For $\sigma_0=10$, $\Gamma\beta$ at the early phase of the expansion is shown in the inset.
\label{fig:FS}}
\end{figure}

Until the deceleration radius, the forward shock can be approximated to freely expand with constant speed. Beyond $R_{\rm dec}$, the forward shock evolves adiabatically with decaying $\Gamma \beta$ approximated by the Blandford-McKee (BM) solution \citep{1995ApJ...455L.143S,2005ApJ...628..315Z,2008A&A...478..747G}. The BM solution is a self-similar solution for an ultra-relativistic outflow from a point-like explosion. According to this solution, the Lorentz factor just behind the shock front $\Gamma_{\rm FS}$ evolves as $\Gamma_{\rm FS} \propto t^{-3/2}$ and the downstream profile of $\Gamma$ is expressed as \citep{1976PhFl...19.1130B}
\begin{equation}
    \Gamma=\Gamma_{\rm FS}\left[1+16\Gamma^2_{\rm FS}\left(1-\frac{r}{R_{\rm FS}}\right) \right]^{-1/2},
    \label{eq:BM}
\end{equation}
where $R_{\rm FS}$ is the radius of the forward shock. Our purpose in this section is to check the asymptotic behaviour of our results at the late phase of expansion and discuss the difference in the case of the ejecta with a finite width.

\begin{figure}
\includegraphics[width=\columnwidth]{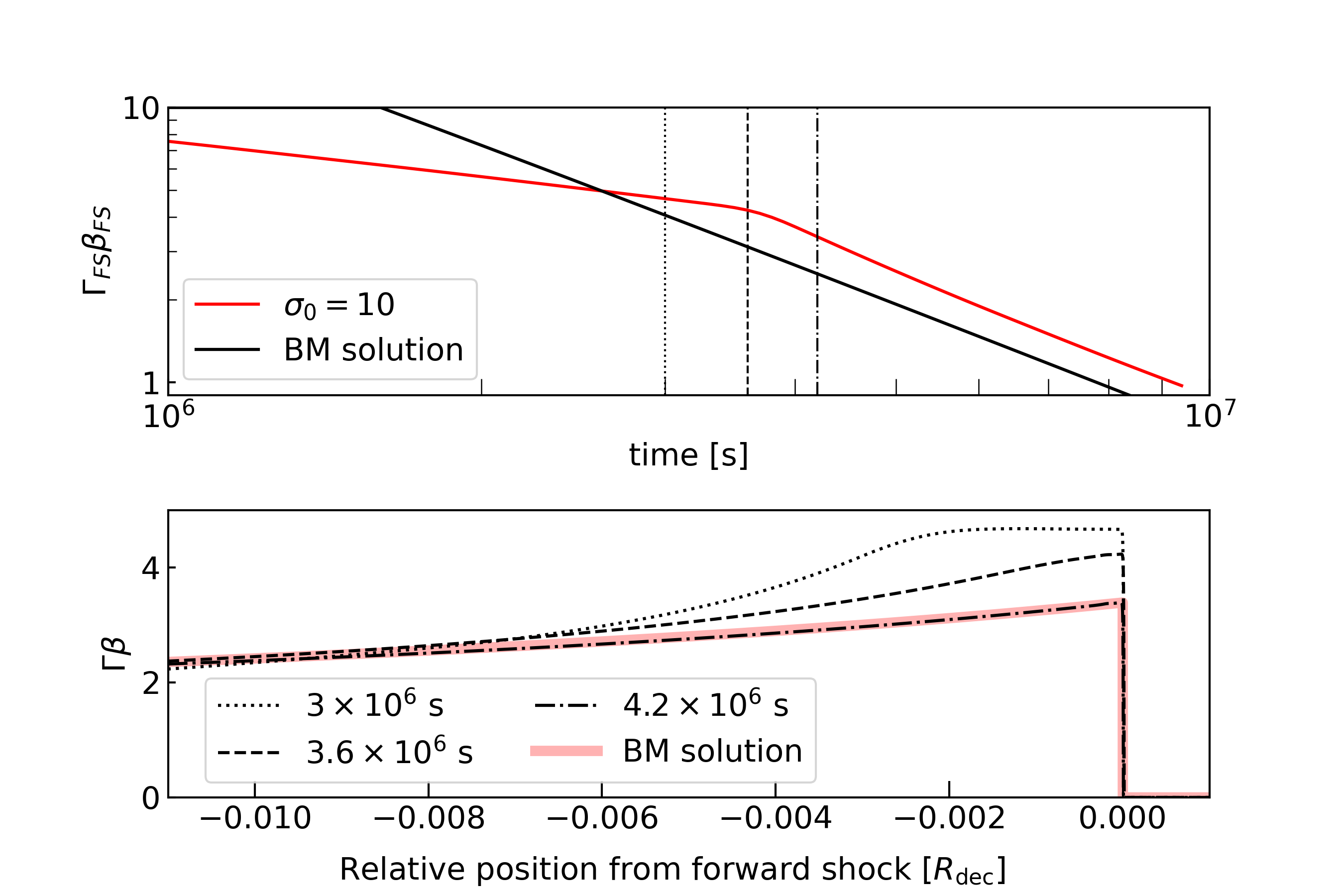}
\caption{Top: The time evolution of the Lorentz factor at the forward shock for the case of $\sigma_0=10$. The vertical lines represent the time before, just at, and after the break time $t_{\rm FS}$. Bottom: The Lorentz factor profiles around the forward shock. The different line types correspond to the same times for the same type of vertical line in the top panel. The red line shows the BM solution described as Eq. (\ref{eq:BM}). 
\label{fig:FScross}}
\end{figure}

Figure \ref{fig:FS} shows the time evolution of $\Gamma\beta$ of the forward shock, which is estimated at the radius where the gas pressure becomes maximum behind the shock front. The black line shows the frequently used approximation for the forward shock 4-velocity, which expresses the transition from the free expansion to the BM solution at the deceleration time $t_{\rm dec}$. Because the mass increase of the shocked external medium leads to the gradual deceleration of the forward shock, a clear break due to the transition from the free expansion to the BM phase does not appear at $t=t_{\rm dec}$. However, Figure \ref{fig:FS} shows that late breaks of $\Gamma\beta$ appear at $t_{\rm FS}\simeq4\times10^6$ s, which is almost independent of the initial magnetisation of the ejecta. As shown in the bottom panel of Figure \ref{fig:FScross}, before the break time $t<t_{\rm FS}$, the profile of $\Gamma\beta$ is flatter than the BM solution just behind the forward shock because of the effect of the initial condition. This flat structure is eliminated gradually as the rarefaction wave R2 approaches the forward shock front. At the break time $t=t_{\rm FS}$, the rarefaction wave just reaches the forward shock front. After the break time $t>t_{\rm FS}$, the profile of $\Gamma\beta$ is almost the same as the BM solution Eq.(\ref{eq:BM}). This break of the $\Gamma\beta$-evolution may affect lightcurves in GRB afterglows.

\subsection{Magnetic energy conversion\label{subsec:energy}}

\begin{figure*}
\includegraphics[width=16 cm]{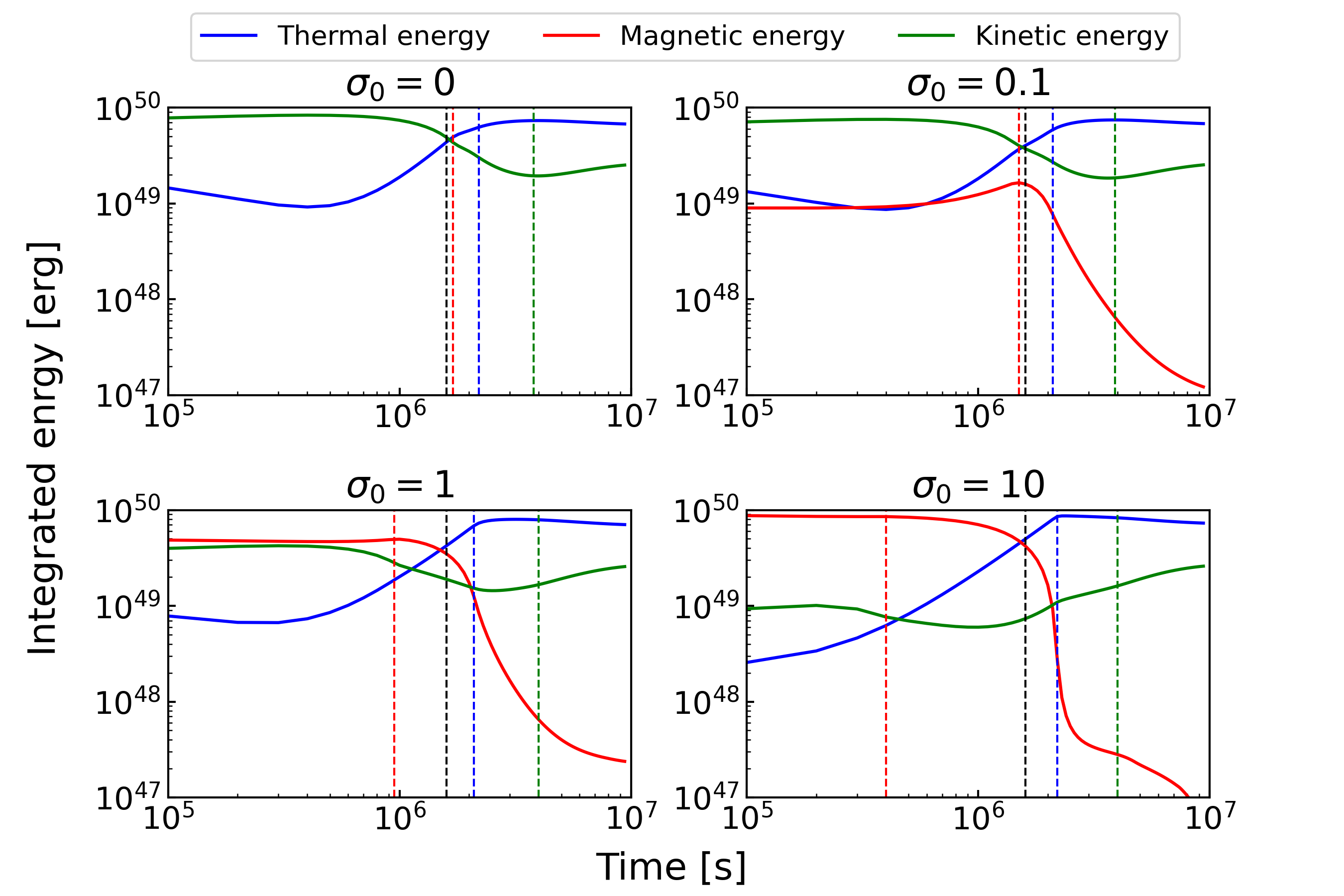}
\caption{The time evolution of the energy in different components of the outflow for different $\sigma_0$. The vertical dashed lines show the analytical deceleration time $t_{\rm dec}$ (black), the reverse shock crossing time $t_{\Delta}$ (red), the R2 crossing time at the CD $t_{\rm CD}$ (blue), and the break time $t_{\rm FS}$ (the R2 crossing time at the FS, green). The blue, red, and green solid lines show the thermal, magnetic, and kinetic energies, respectively.
\label{fig:E}}
\end{figure*}

The time evolutions of the different energy components are shown in Figure \ref{fig:E}. We calculate the kinetic energy component $E_{\rm kin}$, the thermal energy component $E_{\rm th}$, and the magnetic energy component $E_{\rm mag}$ from Eq. (\ref{eq:energy}) as,
\begin{equation}
E_{\rm kin}\equiv\int 4\pi r^2dr\left[\rho c^2\Gamma(\Gamma-1)\right],
\end{equation}
\begin{equation}
E_{\rm th}\equiv\int 4\pi r^2dr\left[(\epsilon+p-\rho c^2)\Gamma^2-p)\right],
\end{equation}
\begin{equation}
E_{\rm mag}\equiv\int 4\pi r^2dr\left[\frac{B^2}{4\pi}-\frac{B^2}{8\pi\Gamma^2}\right],
\end{equation}
respectively.
There are 4 characteristic times as shown in Figure \ref{fig:E}: the deceleration time $t_{\rm dec}$ defined with Eq. (\ref{eq:rdec}), the reverse shock crossing time $t_\Delta$, the break time $t_{\rm FS}$, and the R2 crossing time at the CD $t_{\rm CD}$ defined later.

For $\sigma_0=0$, at the analytical deceleration time $t_{\rm dec}$, the kinetic energy and the thermal energy are roughly in equipartition. The maximum thermal energy is achieved at around $4\times10^6$ s, which corresponds to the break time $t_{\rm FS}$ (the RS crossing time at the FS). After that, the thermal energy gradually decreases by adiabatic expansion. 

For $\sigma_0=0.1$ and $\sigma_0=1$, the thermal and the kinetic energy evolutions are similar to the case of $\sigma_0=0$. The magnetic energy increases by the reverse shock compression until the reverse shock crossing time $t_\Delta$. At the deceleration time $t_{\rm dec}$, the thermal energy is almost equipartition with the kinetic energy for $\sigma_0=0.1$, while with the magnetic energy for $\sigma_0=1$. This means that almost half of the initial energy of the ejecta is converted to the thermal energy of the shocked external medium at $t_{\rm dec}$. For $\sigma_0=0.1$, the maximum thermal energy is achieved at $t_{\rm FS}$. While for $\sigma_0=1$, the thermal energy reaches maximum slightly before $t_{\rm FS}$. After that, the adiabatic expansion reduces the thermal energy gradually.

\begin{figure}
\includegraphics[width=\columnwidth]{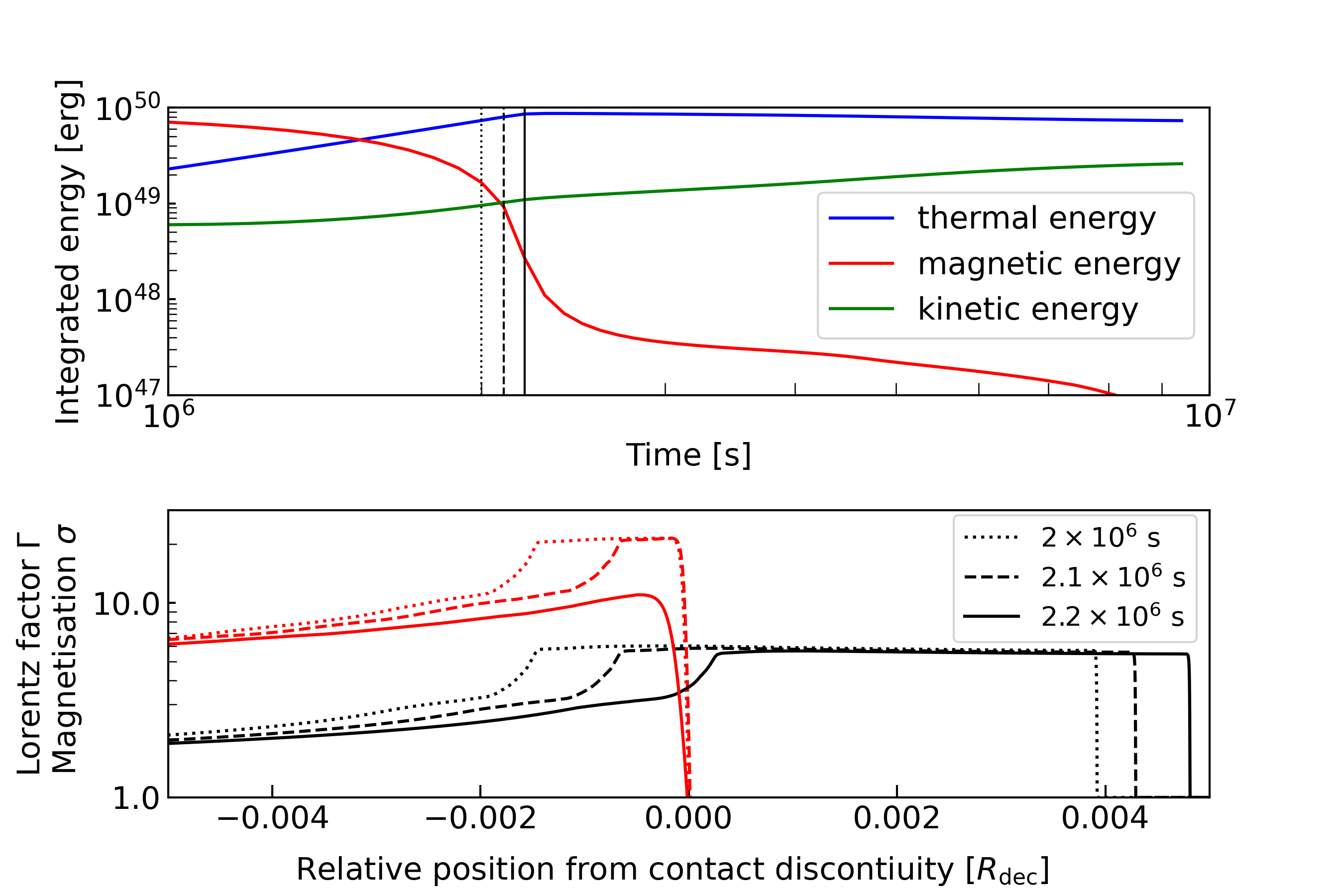}
\caption{Top: The time evolutions of the energy components for $\sigma_0=10$. The vertical lines represent the time before, and just at the rarefaction wave R2 crossing time $t_{\rm CD}$. The coloured lines show the same ones as Figure \ref{fig:E}. Bottom: The Lorentz factor and magnetisation profiles around the contact discontinuity. The black lines show $\Gamma$ and the red lines show $\sigma$. The different types of lines correspond to the same times for the same type of vertical line in the top panel.
\label{fig:CDcross}}
\end{figure}

For $\sigma_0=10$, as the reverse shock is weak due to the high $\sigma$, the magnetic energy is almost constant in the initial phase. At the deceleration time $t_{\rm dec}$, the magnetic energy and the thermal energy are in equipartition. As shown in Figure \ref{fig:CDcross}, as the rarefaction wave R2 approaches the contact discontinuity, the flat structure of $\sigma$ in the ejecta disappears. We define the time at which the R2 crosses the CD as $t_{\rm CD}$. At $t_{\rm dec}\leqq t\leqq t_{\rm CD}$, only the magnetic energy decreases drastically, while the kinetic and the thermal energy increase. This implies that the magnetic energy is converted into the kinetic energy of the shocked external medium by the magnetic pressure, and a part of that energy is dissipated into the thermal energy by the forward shock \citep{2011MNRAS.411.1323G,2012MNRAS.422..326K}. The maximum thermal energy is achieved at $t_{\rm CD}$. After that, the adiabatic expansion reduces the thermal energy gradually.

In GRB afterglow models, the deceleration time $t_{\rm dec}$ is assumed as the peak emission time from the forward shock \citep{2005ApJ...628..315Z}. We can estimate the initial $\Gamma$ from this peak time. However, for the thick shell cases, our simulation implies that the emission has peaked at $t_{\rm FS}$ for $\sigma_0<1$ and at $t_{\rm CD}$ for $\sigma_0>1$. The difference in the magnetisation may affect the emission from not only the reverse shock but also the forward shock. Since the reverse shock dynamics might be suffered from the Rayliegh-Taylor instability at the contact discontinuity as discussed in \citet{2013ApJ...775...87D} by their 2D simulation, we will further study this impact on magnetic energy dissipation by our 2D simulations in future.

\section{Multiple magnetic bullets and winds}\label{sec:blazar}

\begin{figure*}
\includegraphics[width=16 cm]{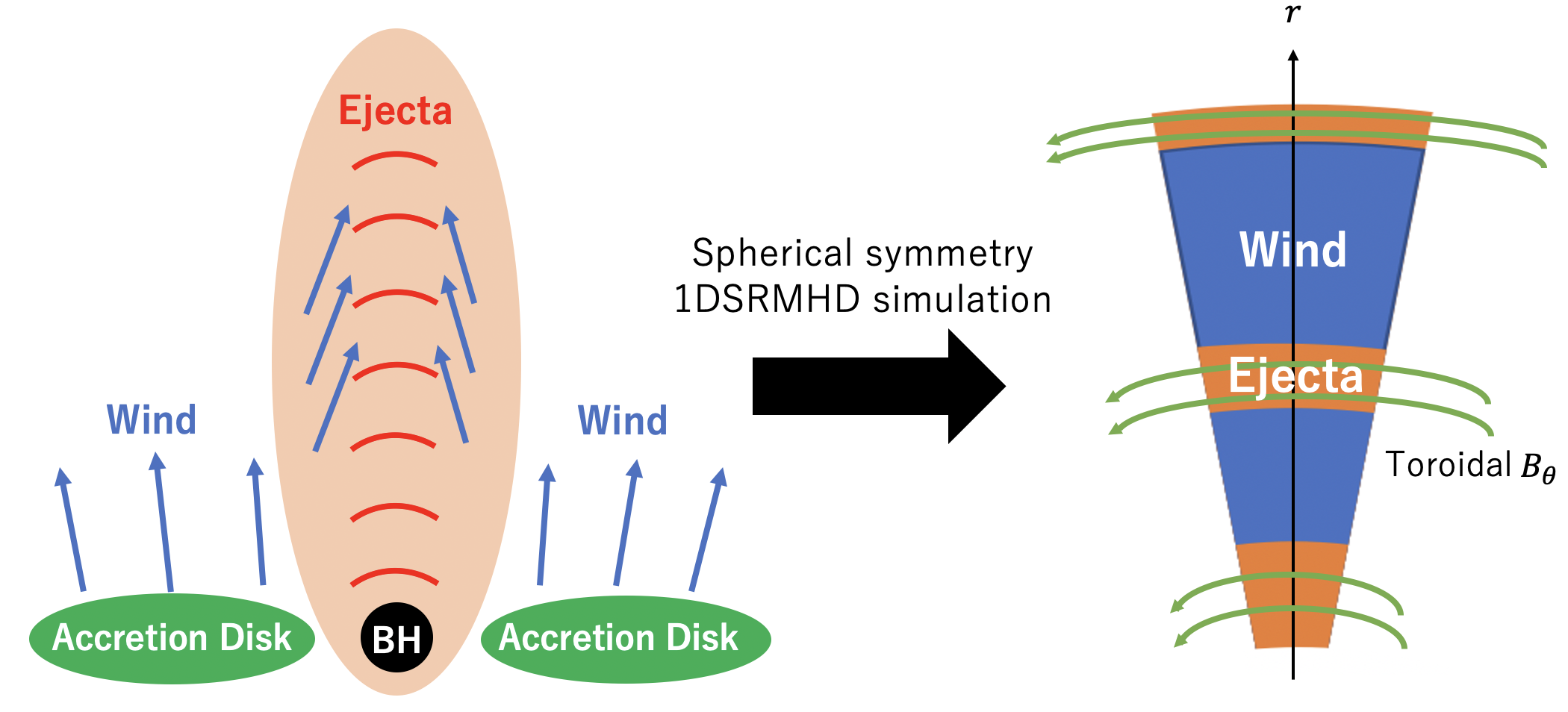}
\caption{Schematic picture of the multiple magnetic bullets model in Section \ref{sec:blazar}. 
\label{fig:situation}}
\end{figure*}

In this section, we demonstrate the energy dissipation of multiple relativistic magnetic bullets interacting with low-magnetised winds as shown in Figure \ref{fig:situation}. The situation is close to the internal shock model, which was originally proposed to explain Blazar flares and GRB prompt emissions \citep{1978MNRAS.184P..61R,1994ApJ...430L..93R,1997ApJ...490...92K,1998MNRAS.296..275D}. Differently from the internal shock model, however, we consider non-relativistic winds from an accretion disk between the relativistic ejecta (magnetic bullets). We assume that the ejecta is collimated by the ram pressure of the winds. At inner radii, the centrifugal barriers may prevent the winds from invading between the ejecta. But at outer radii, the surrounding medium confines the winds so that the winds may penetrate low-pressure regions between the ejecta. We carry out 1D spherically symmetrical SRMHD simulations. We inject ejecta and winds into the simulation box randomly and alternately as shown in the right-side panel of Figure \ref{fig:situation}. 

As studied in the previous section, a collision between an ejecta and a wind can lead to efficient shock dissipation. A part of the shock dissipated energy can be converted into non-thermal electron energy, which will be released as radiation energy from low-magnetised regions. 

\subsection{Initial setup}

The simulation box size is $[R_0,20R_0]$, where $R_0=10^{16}\ \rm{cm}$. The inner boundary is set to be an injection boundary. The outer boundary is set to be an open boundary. The number of static cells is $2\times10^5$, effectively $3.2\times10^6$ via the AMR procedure (see Appendix \ref{Appendix A}). For the initial condition, wind fills the simulation box with the density profile of $\rho\propto r^{-2}$.

We inject ejecta and winds alternately from the inner boundary. The Lorentz factor of the ejecta $\Gamma_{\rm jet}$ are randomly following a cutoff-Gaussian probability distribution: a Gaussian with maximum and minimum cutoffs. The typical Lorentz factor of ejecta is around 10 for Blazars \citep{1993ApJ...407...65G} (GRB jets have so much higher value that simulations for such ultra-relativistic outflows are difficult). Thus, we choose the mean value of $\Gamma_{\rm jet}$ is 10 and its dispersion is 5. The cutoffs are at $10\pm5$. We define the minimum injection duration of the ejecta as
\begin{equation}
t_0=\frac{R_0}{\Gamma_{\rm jet}^2c}.
\end{equation}
The injecting duration of the ejecta $t_{\rm jet}$ is randomly determined with a cutoff-Gaussian probability distribution with the mean value of $3t_0$, the dispersion of $2t_0$, and the cutoffs at $3t_0\pm2t_0$. The initial magnetisation and temperature of the ejecta are fixed as $\sigma_{\rm jet}=10$ and 100 MeV, respectively. The total luminosity of the ejecta is fixed as $L_{\rm jet}=10^{45}$ erg $\rm{s}^{-1}$, which means that the density changes according to changing $\Gamma_{\rm jet}$.

\begin{table}
 \caption{Parameters and results of the simulations in Section \ref{sec:blazar}}
 \label{table:models}
 \centering
  \begin{tabular}{ccccc}
   \hline
   Model & $\eta$ & $t_{\rm wind}$ & $f_{\rm max}$ & Behaviour \\
   \hline \hline
   A & $10^{-4}$ & $(1\pm0.1)t_w$ & 0.15 & stationary \\
   B & $10^{-2}$ & $(1\pm0.1)t_w$ & 0.07 & flare-like \\
   \hline
   C & $10^{-3}$ & $(1\pm0.1)t_w$ & 0.12 & both \\
   D & $10^{-3}$ & $(1\pm0.1)t_w/2$ & 0.075 & stationary \\
   \hline
  \end{tabular}
\end{table}

For winds, the velocity, magnetisation, and temperature are fixed as $0.1c$, $\sigma_{\rm wind}=10^{-10}$, and 100 MeV, respectively. To reduce the computational cost due to AMR refinement procedures, we assume the mean injection duration of winds $t_w$ is significantly longer than $t_0$. The injecting duration of the winds $t_{\rm wind}$ is determined by a cutoff-Gaussian probability distribution with the mean value of $t_w=R_0/c$, the dispersion of $0.1t_w$, and cutoffs at $t_w\pm0.1t_w$.  

The total luminosity of the winds $L_{\rm wind}$ is fixed in a simulation run. We parameterise $L_{\rm wind}$ as
\begin{equation}
   \eta\equiv\frac{L_{\rm{wind}}}{L_{\rm jet}}.
\end{equation}
If the ejecta energy is totally transferred into the swept-up wind energy, we can write
\begin{equation}
L_{\rm jet}t_{\rm jet}=\Gamma_{\rm jet}^2L_{\rm wind}t_{\rm wind}.
\end{equation}
For $t_{\rm jet}=3t_0$, $t_{\rm wind}=t_w$ and $\Gamma_{\rm jet}=10$, we obtain
\begin{equation}
\frac{L_{\rm wind}}{L_{\rm jet}}=\frac{t_{\rm jet}}{\Gamma^2_{\rm jet}t_{\rm wind}}=3\times10^{-4}\equiv\eta_{\rm crit}.
\end{equation}
If $\eta$ is much smaller than this critical value $\eta_{\rm crit}$, the wind luminosity may be not significant to dissipate the energy of the magnetic bullet. Conversely, we expect that most of the ejecta energy can be transferred into the energy of the shocked wind for $\eta>\eta_{\rm crit}$. To investigate the $\eta$-dependence on the outflow structure and the dissipation efficiency, we carry out four simulations listed in Table \ref{table:models}. We probe the difference due to different $\eta$ with models A and B, and the different $t_{\rm wind}$ with models C and D. 

\subsection{Outflow structures}
\label{sec:4.2}

\begin{figure*}
\includegraphics[width=16 cm]{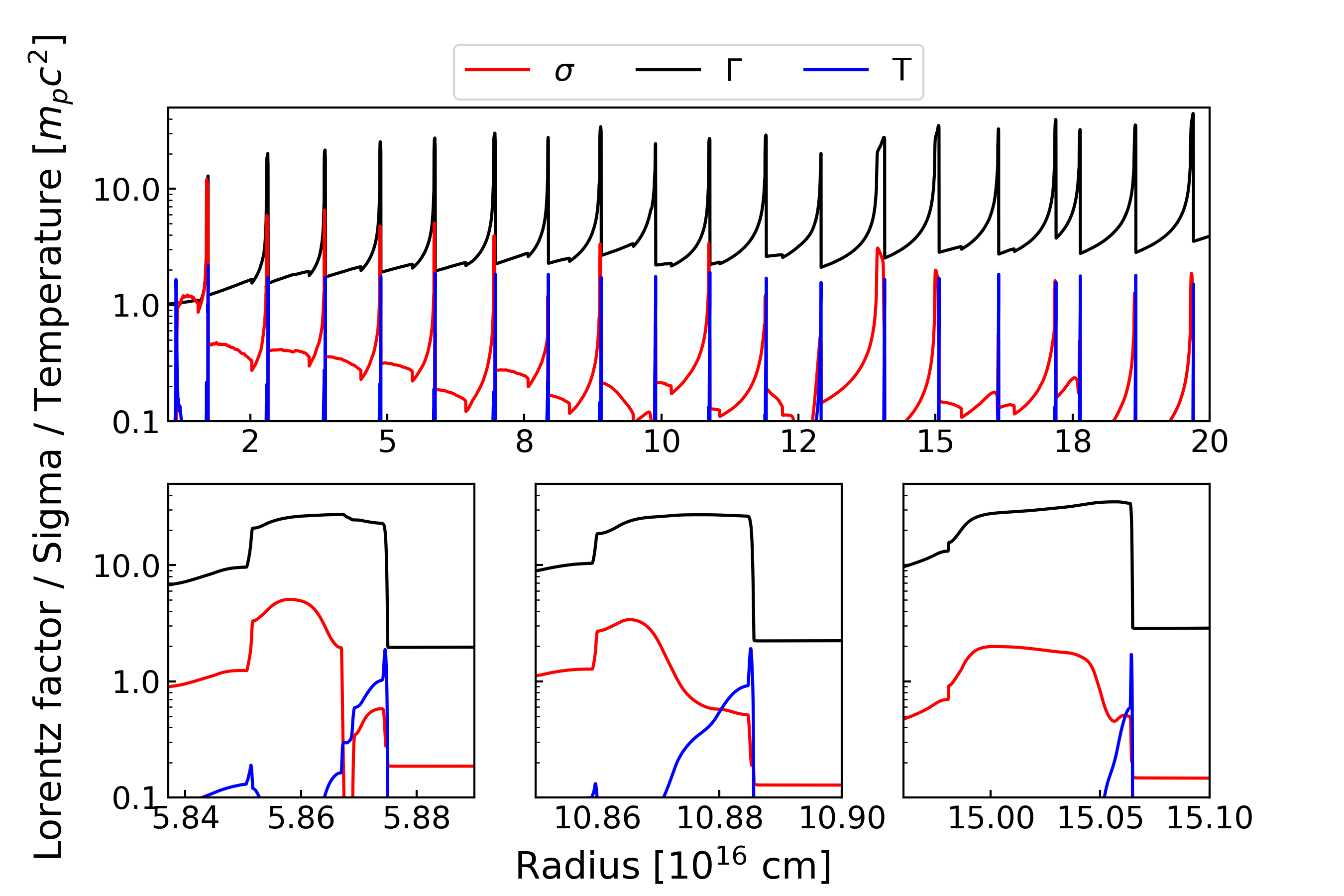}
\caption{A snapshot at $2\times10^6$ s for model A. The red, black, and blue line shows the magnetisation $\sigma$, the Lorentz factor $\Gamma$, and the proton temperature $T$, respectively. Top panel: Whole simulation box. Bottom panels: The enlarged view of high $\Gamma$ regions.
\label{fig:SS-4}}
\end{figure*}

\begin{figure*}
\includegraphics[width=16 cm]{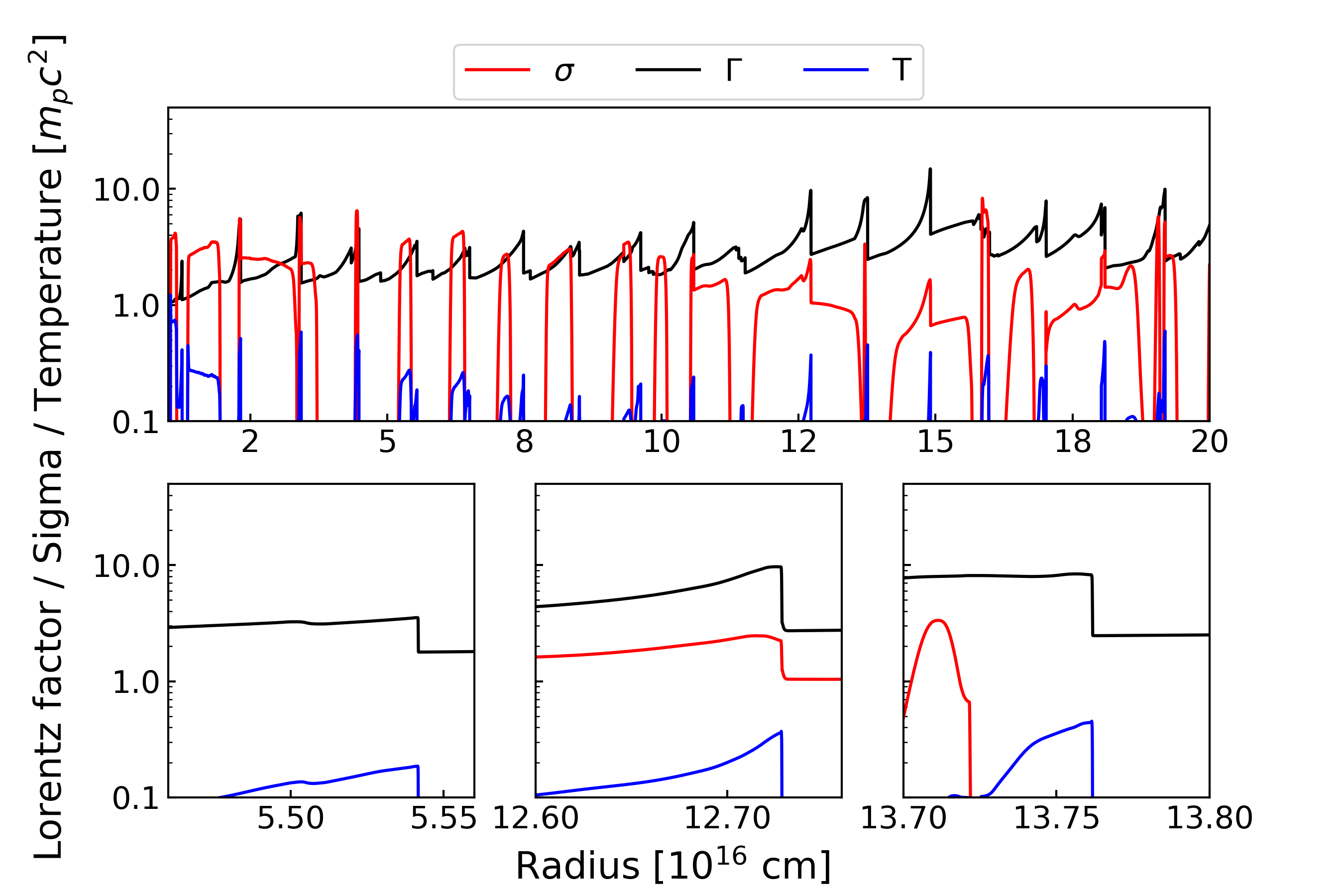}
\caption{Same as Figure \ref{fig:SS-4} but for model B.
\label{fig:SS-2}}
\end{figure*}

Snapshots of the outflows at $2\times10^6$ s ($\sim10^3t_0$) are shown in Figure \ref{fig:SS-4}, \ref{fig:SS-2}, \ref{fig:SS-3}, and \ref{fig:SS-3(short)} for models A, B, C, and D, respectively. The influence of the initial condition has already disappeared at this stage so the simulations are in quasi-steady states. High-temperature regions are the sites where energy dissipation occurs. Almost all of the efficient dissipation regions are concentrated around the discontinuities of the Lorentz factor. This means that efficient shock dissipation occurs through interactions between ejecta and low-magnetised plasma. Because the outflow structure is dependent on wind models, we explain the difference between models A and B as follows (see Appendix \ref{Appendix B} for models C and D). 

Figure \ref{fig:SS-4} shows the outflow structure for model A ($\eta=10^{-4}<\eta_{\rm crit}$). The high-$\sigma$ ejecta between the low-pressure winds is gradually accelerated. This process is equivalent to the impulsive acceleration discussed in \citet{2011MNRAS.411.1323G} \citep[see also][]{2012MNRAS.422..326K}. The Lorentz factor of the front edge of the ejecta increases as $\Gamma\sim\Gamma_{\rm jet}(R/R_0)^{1/3}$ until the rarefaction wave from the rear edge will catch up with it. Behind the ejecta, there exist long rarefaction tails with moderate magnetisation. The ram pressure of the winds is weak so the winds are compressed by the rarefaction tails. 

The bottom left panel of Figure \ref{fig:SS-4} shows the interaction between the head of an inner ejecta ($<5.87\times10^{16}$ cm), a wind (the narrow region between high-$\sigma$ regions), and the rear tail of an outer ejecta ($>5.87\times10^{16}$ cm). The high $\Gamma$ and temperature of the wind indicate that the magnetic energy of the inner ejecta is transferred into the kinetic and thermal energies of the wind (see Section \ref{sec:GRB} in detail). The outer jump of $\Gamma$ corresponds to the shock front propagating in the rarefaction tail of the outer ejecta. The apparent low temperature in the shocked wind region is due to thermal pressure acceleration, which transfers the thermal energy of the shocked wind into the kinetic and the thermal energy of the rarefaction tail of the outer ejecta. As shown in the bottom middle and right panels of Figure \ref{fig:SS-4}, the wind components are squashed at larger radii, so that they are below the resolution in our simulations.

In model A, the shocked rarefaction tails, where $\sigma<1$ and $\Gamma>10$, are regarded as kinetically dominated relativistic hot outflows, which can be efficient emission sites. 

On the other hand, Figure \ref{fig:SS-2} shows a very different outflow structure for model B ($\eta=10^{-2}>\eta_{\rm crit}$). It shows that the Lorentz factor of the ejecta is no larger than the value at injection. This is due to the heavy mass of the winds. The interaction between the ejecta and winds leads to a significant dissipation of the kinetic energy of the ejecta. A fraction of ejecta at outer radii has a high Lorentz factor due to impulsive acceleration. Differently from model A, the $\sigma$ values of the rarefaction tails are mostly higher than one. This is because the heavy winds prohibit rarefaction tails from expanding, which reduces their magnetisation. 

The bottom left panel of Figure \ref{fig:SS-2} shows a shock in the wind. Because the shock is mildly relativistic, the temperature is lower than the cases in model A. In such shocked regions, the emission efficiency may not be so high. The shock waves in the winds tend to catch up the rarefaction tail of the preceding ejecta at inner radii. Such shocks propagate in the rarefaction tail eventually. One of such shocks is shown in the bottom middle panel of Figure \ref{fig:SS-2}. Since the strong ram pressure of the winds prevents the preceding rarefaction tail from expanding enough to reduce its magnetisation, the shocks are weaker than those in model A, leading to less energy dissipation. However, we find succeeding acceleration of such shocks by magnetic pressure even in the rarefaction tails. Some of such shocks collide with the preceding wind again (see the bottom right panel of Figure \ref{fig:SS-2}), which leads to the high efficiency of the energy dissipation at outer radii. Thus, kinetically-dominated relativistic hot outflows tend to be generated at outer radii for model B. We will further discuss this radial dependency in the following section.

\subsection{Dissipation efficiency}

\begin{figure*}
\includegraphics[width=16 cm]{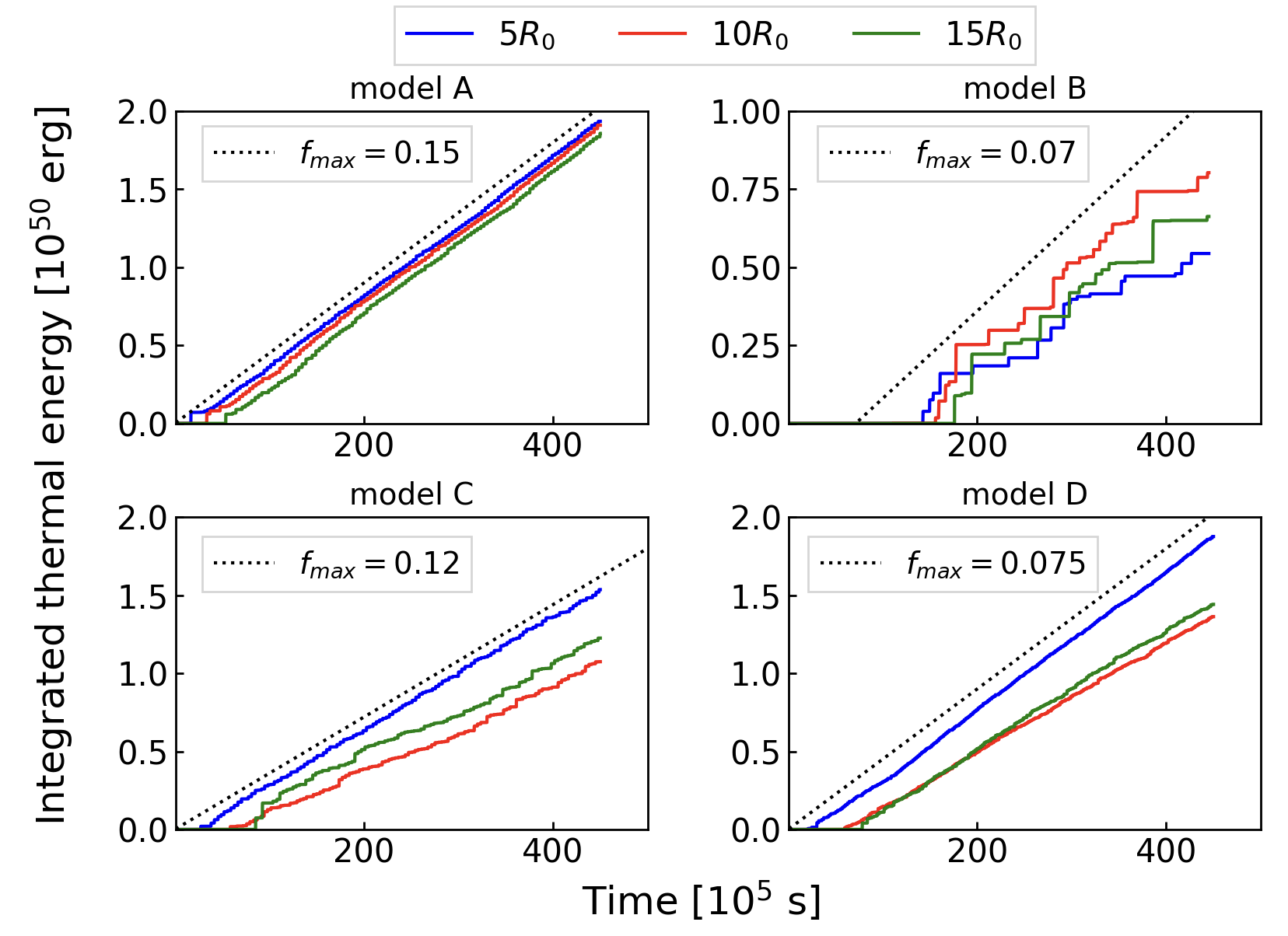}
\caption{The time evolution of $E_{\rm th}(R,\ t)$ at $R=5,10,15R_0$. Each panel corresponds to a different wind model. Each coloured line shows the integrated thermal energy at the different radii. The black dotted lines are $f_{\rm max}L_{\rm in}t$. 
\label{fig:luminosity_average}}
\end{figure*}

\begin{figure}
\includegraphics[width=\columnwidth]{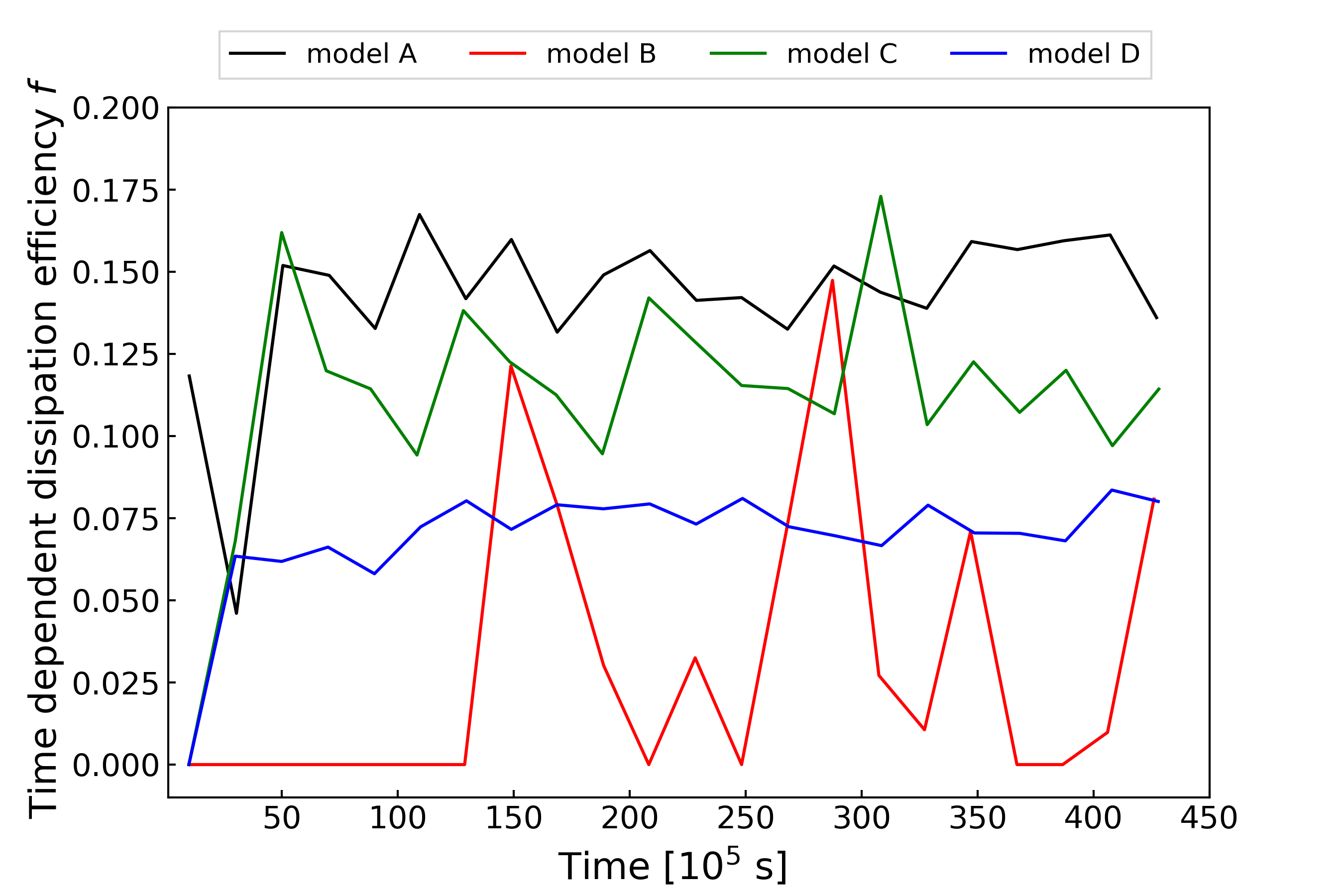}
\caption{The time dependent dissipation efficiency $f$ at $5\times10^{16}$ cm as a function of time.
\label{fig:efficiency}}
\end{figure}

In the previous section, we find that interactions between magnetised ejecta and weakly magnetised plasma (winds or rarefaction tails) can produce kinetically dominated relativistic hot outflows. In this section, we estimate the magnetic dissipation efficiency. We calculate the thermal luminosity of outflows passing some radius $R$ as
\begin{equation}
L_{\rm th}=4\pi R^2\left[(\epsilon+p_g-\rho c^2)\Gamma^2c\beta\right].
\end{equation}
Because our interest is the thermal energy of the kinetically dominated relativistic outflows, we estimate the thermal energy for only outflows with $\sigma<1$ and $\Gamma>5$ as
\begin{equation}
    E_{\rm th}(R,\ t)=\int_0^tdt'L_{\rm th}(R,\ t';\ \sigma<1,\ \Gamma>5).
\end{equation}
The result is shown in Figure \ref{fig:luminosity_average}. In all cases, the integrated thermal energy increases almost as a monotonic function of time. The maximum average dissipation efficiency $f_{\rm max}$ corresponds to the slope of the dotted lines in Figure \ref{fig:luminosity_average} is calculated by
\begin{equation}
    \frac{dE_{\rm th}}{dt}=f_{\rm max}L_{\rm in},
\end{equation}
where $L_{\rm in}$ is the time-averaged total injection luminosity of the outflow. Without the condition of $\sigma<1$ and $\Gamma>5$, we find that the efficiency is about 18\% for models A, B and C, while 14 \% for model D. Irrespective of the high $\sigma$ values at injection, significantly high efficiencies are successfully obtained in our simulations.

For model A, the integrated energy increases smoothly with an efficiency of 15\% independent of the radius. This is because almost all of the dissipation occurs through interactions between ejecta and rarefaction tails at any radius. On the other hand, model B shows that dissipation at $R=10R_0$ is more effective than that at $R=5R_0$. Because most outflows are mildly relativistic ($\Gamma<5$) as shown in Figure \ref{fig:SS-2}, the integrated energy increases step-wise at the times when a high $\Gamma$ outflow arrives. As we pointed out in Section \ref{sec:4.2}, high $\Gamma$ shocks in the winds tend to be produced after experiencing magnetic pressure acceleration in the preceding rarefaction tails at outer radii.

For models C and D (intermediate case for $\eta$), the average dissipation efficiency becomes larger at the inner radii ($R=5R_0$). The wind regions are below our numerical resolution at outer radii (see Appendix \ref{Appendix B}), so low magnetised outflows exist only at inner radii, where the magnetic energy can be efficiently dissipated. The maximum dissipation efficiency of model C is higher than that of model D. This is because a short wind duration suppresses the energy dissipation via the interaction with the winds.

To clarify the behaviour of the dissipation, we estimate the time-dependent dissipation efficiency at $R=5\times10^{16}$ cm using the following definition
\begin{equation}
    f(t)=\frac{\int_t^{t+\delta t} dt'L_{\rm th}(R,\ t';\ \sigma<1,\ \Gamma>5)}{L_{\rm in}\delta t},
\end{equation}
where $\delta t$ is set to be sufficiently long as $2\times10^6$s$\sim 10^3 t_0$ to produce a relatively smoothed shape of $f$. 
The results are plotted in Figure \ref{fig:efficiency}. After reaching the quasi-steady state ($\sim10^7$ s), the time-dependent dissipation efficiency is almost stationary for model A (13 $\sim$ 17\%) and D (7 $\sim$ 8\%). On the other hand, it fluctuates significantly for model B (0 $\sim$ 15\%). The behaviour in model C is intermediate (10 $\sim$ 17\%). In spite of the low average dissipation efficiency for model B, the flare-like dissipation may be favourable to explain the gamma-ray flares frequently seen in blazar observations.

\section{Summary} \label{sec:dis}

We demonstrate the energy dissipation of the magnetic field in relativistic outflows by 1D ideal MHD simulations.
In the external shock case, where a single ejecta interacts with an external medium, we have confirmed the efficient energy transfer from the magnetised ejecta to the shocked external medium for a wide range of the magnetisation parameter as $\sigma=0$--$10$. The deceleration time $t_{\rm dec}$ is a good indicative parameter for the energy equipartition time, but the actual transition to the Blandford-McKee phase occurs later when the rarefaction wave catches up with the forward shock front. In high $\sigma$ cases, the energy dissipation into the shocked medium occurs when the rarefaction tail catches up with the contact discontinuity. This energy dissipation time is significantly later than the reverse shock crossing time.

We also conducted numerical experiments of the magnetic energy dissipation of multiple ejecta launched intermittently. Between the relativistic magnetised ejecta, we inject non-relativistic winds without a magnetic field.
Kinetically dominated relativistic outflows with a significantly high temperature can be produced by interactions between strongly magnetised relativistic ejecta with weakly magnetised winds or rarefaction tails of other ejecta.
We found that the energy dissipation efficiency into hot, low-magnetised, and relativistic outflows is about 10 \%.
When the luminosity of the winds is relatively low, the dissipation is mainly due to the interaction with the rarefaction tails. The typical magnetisation in the dissipation regions is $\sigma\sim 0.1$ in those cases. When the wind luminosity is strong ($\eta>\eta_{\rm crit}$), the magnetic energy is efficiently converted into the thermal energy in the winds, and the magnetisation in the dissipation regions can be $\sigma \ll 1$. However, the Lorentz factors of the dissipation regions tend to be low. While the energy dissipation is stationary for low wind luminosity, a flare-like dissipation occurs for high wind luminosity.

\section*{Acknowledgements}

The authors are grateful to K. Kawaguchi and an anonymous referee for thoughtful discussions and suggestions for improvements to this work. The authors thankfully acknowledge the computer resources provided by the Institute for Cosmic Ray Research (ICRR), the University of Tokyo. 
This work is supported by the joint research program of ICRR, and JSPS KAKENHI Grant Numbers JP23KJ0692 (Y.K.), JP22K03684, JP23H04899 (K.A.),  JP22K14032 (T.O.), and JP18K13594, JP23K03448 (T.K.). 


\section*{Data Availability}

The data underlying this article will be shared on reasonable request to the corresponding author.



\bibliographystyle{mnras}
\bibliography{main} 




\appendix

\section{Adaptive mesh refinement} \label{Appendix A}

\begin{figure}
\includegraphics[width=\columnwidth]{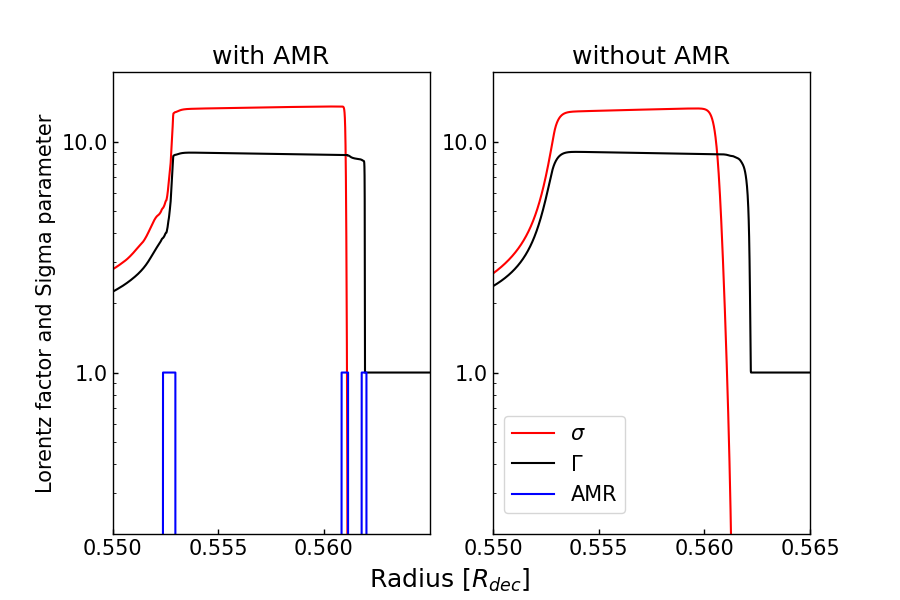}
\caption{Comparison with and without AMR. The red and black lines show $\sigma$ and $\Gamma$, respectively. The blue enclosed region reveals the AMR region.
\label{fig:AMR}}
\end{figure}

Adaptive mesh refinement (AMR) is a method of achieving higher accuracy of a solution within limited regions of simulation. In our simulation code, we use AMR mesh around all the discontinuities of the magnetic pressure. If the magnetic pressure of a cell becomes 1.05 times higher than its neighbour cells in each simulation time step, we divide 20 static cells around that cell into $2^4$ for each static cell. Figure \ref{fig:AMR} shows the comparison with and without AMR cells.

\section{Outflow structures for different wind duration} \label{Appendix B}

\begin{figure}
\includegraphics[width=\columnwidth]{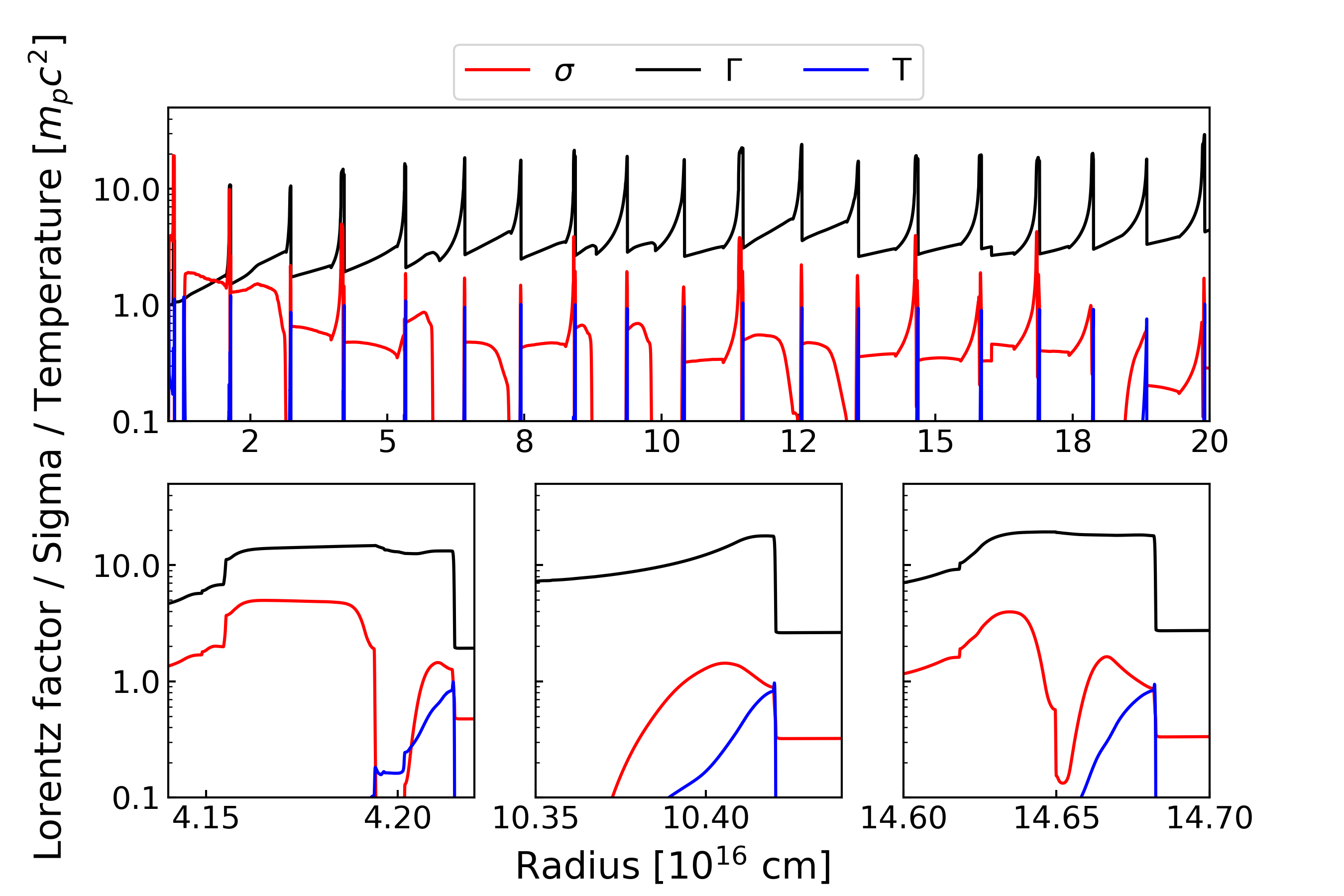}
\caption{Same as Figure \ref{fig:SS-4} for model C.
\label{fig:SS-3}}
\end{figure}

\begin{figure}
\includegraphics[width=\columnwidth]{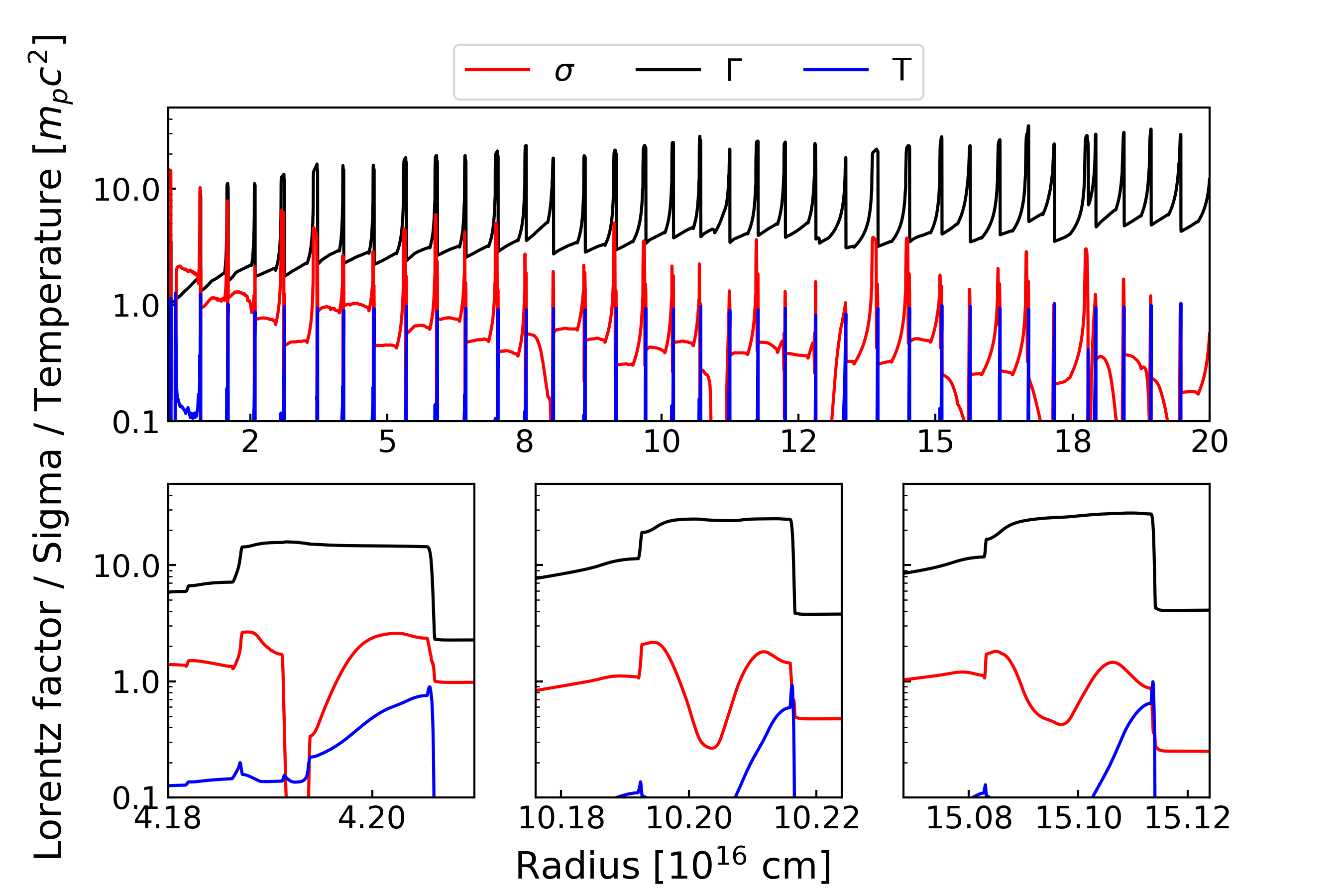}
\caption{Same as Figure \ref{fig:SS-4} for model D.
\label{fig:SS-3(short)}}
\end{figure}

Figure \ref{fig:SS-3} shows the outflow structure for model C ($\eta=10^{-3}$). The Lorentz factor increases to around 20. The magnetisation in the rarefaction tails is slightly less than 1. There remain some wind regions (identified with $\sigma=0$) resolved with our simulation even at the larger radius compared to model A (see Figure \ref{fig:SS-4}), but not so wide as those in model B (see Figure \ref{fig:SS-2}). The interaction with both the winds and the rarefaction tails produces thermal energy through the magnetic energy conversion mechanism. At larger radii, the wind widths are thinner than the numerical resolution, leading to a less dissipation efficiency there.

Figure \ref{fig:SS-3(short)} shows the outflow structure for model D, where the time interval between ejecta is shorter than that in model C. Compared with model C, the Lorentz factor increases to around 30, and most of the wind regions are below our resolution. Because of the short duration of the winds, the critical value $\eta_{\rm crit}$ becomes larger as $(\eta_{\rm crit}^{\rm model\ D}=6\times10^{-4}>\eta_{\rm crit})$. Thus, the outflow structure is similar to that in model A, except for the high magnetisation in the rarefaction tails. The expansion of the rarefaction tails is not enough to reduce $\sigma$ due to the short interval. Therefore, a shorter duration of the winds leads to less magnetic energy dissipation.


\bsp	
\label{lastpage}
\end{document}